\RequirePackage{etoolbox}
\csdef{input@path}{{style/}{graphics/}}
\documentclass[aoas,MSNbibl,nameyear,secthm,seceqn,dvips]{arximspdf}
\makeatletter
   \@ifpackageloaded{graphicx}{}{\usepackage{graphicx}}
\makeatother
\usepackage{multirow}
\usepackage{graphicx}

\doi{10.1214/14-AOAS800}
\volume{9}
\issue{1}
\pubyear{2015}
\firstpage{166}
\lastpage{199}
\docsubty{FLA}

\makeatletter
\newcommand{\ds}{\displaystyle}
\newcommand{\rrvert}{\vert}
\newcommand{\llvert}{\vert}
\makeatother

\begin{document}
\begin{frontmatter}

\title{Estimating network degree distributions under sampling: An
inverse problem, with applications to monitoring social media
networks\thanksref{T1}}
\runtitle{Estimating network degree distributions}

\begin{aug}
\author[A]{\fnms{Yaonan}~\snm{Zhang}\corref{}\thanksref{m1}\ead[label=e1]{yaonanzh@bu.edu}},
\author[A]{\fnms{Eric D.}~\snm{Kolaczyk}\thanksref{m1}\ead[label=e2]{kolaczyk@bu.edu}}
\and
\author[B]{\fnms{Bruce D.}~\snm{Spencer}\thanksref{m2}\ead[label=e3]{bspencer@northwestern.edu}}
\runauthor{Y. Zhang, E. D. Kolaczyk and B. D. Spencer}
\affiliation{Boston University\thanksmark{m1} and Northwestern
University\thanksmark{m2}}
\address[A]{Y. Zhang\\
E. D. Kolaczyk\\
Department of Mathematics and Statistics\\
Boston University\\
Boston, Massachusetts 02215\\
USA\\
\printead{e1}\\
\phantom{E-mail:\ }\printead*{e2}}
\address[B]{B. D. Spencer\\
Department of Statistics\\
Northwestern University\\
Evanston, Illinois 60208\\
USA\\
\printead{e3}}
\end{aug}
\thankstext{T1}{Supported in part by AFOSR award 12RSL042 and NSF Grant
CNS-0905565.}

%
\received{\smonth{5} \syear{2013}}
%
\revised{\smonth{8} \syear{2014}}

\begin{abstract}
Networks are a popular tool for representing elements in a system and
their interconnectedness. Many observed networks can be viewed as only
samples of some true underlying network. Such is frequently the case,
for example, in the monitoring and study of massive, online social
networks. We study the problem of how to estimate the degree
distribution---an object of fundamental interest---of a true
underlying network from its sampled network. In particular, we show
that this problem can be formulated as an inverse problem. Playing a
key role in this formulation is a matrix relating the expectation of
our sampled degree distribution to the true underlying degree
distribution. Under many network sampling designs, this matrix can be
defined entirely in terms of the design and is found to be
ill-conditioned. As a result, our inverse problem frequently is
ill-posed. Accordingly, we offer a constrained, penalized weighted
least-squares approach to solving this problem. A Monte Carlo variant
of Stein's unbiased risk estimation (SURE) is used to select the
penalization parameter. We explore the behavior of our resulting
estimator of network degree distribution in simulation, using a variety
of combinations of network models and sampling regimes. In addition, we
demonstrate the ability of our method to accurately reconstruct the
degree distributions of various sub-communities within online social
networks corresponding to Friendster, Orkut\ and LiveJournal. Overall,
our results show that the true degree distributions from both
homogeneous and inhomogeneous networks can be recovered with
substantially greater accuracy than reflected in the empirical degree
distribution resulting from the original sampling.
\end{abstract}

\end{frontmatter}

\section{Introduction}

Many networks observed or investigated today are samples of much larger
networks [\citet{kolaczyk2009statistical}, Chapter~5]. Let $G=(V,E)$ be
a graph representing a network, with vertex set $V$ and edge set $E$.
Similarly, let $G^{*}=(V^{*},E^{*})$ denote a subgraph of $G$,
representing a part of the network obtained through some sort of
network sampling. Although practitioners typically speak of \textit{the}
network when presenting empirical results, frequently it is only a
sampled version $G^*$ (or some function thereof, such as when sampling
yields estimates of vertex degrees directly) of some true underlying
network $G$ that is available to them, either by default or design. A
central statistical question in such studies, therefore, is how much
the properties of the sampled network reflect those of the true network.

Sampling is of particular interest in the context of online social
networks. One reason for such interest is that these networks are
usually very large. For example, social networks from Friendster,
LiveJournal, Orkut and Amazon have been studied in~\citet
{yang2012defining} having, respectively,
$117.7M, 4.0M, 3.0M$ and $0.33M$ vertices and $2586.1M$, $34.9M$,
$117.2M$ and $0.92M$ edges. Similarly in \citet{ribeiro2010estimating},
networks from Flickr and Youtube were studied having millions of
vertices and edges as well. The large size of these social networks
makes it costly querying the entire network, particularly if the goal
is to monitor these networks regularly over time. In addition, the
decentralized nature of many such networks frequently means that few---if any---people or organizations have complete access to the data.

The topic of network sampling goes back at least to the seminal work of
Ove Frank and his colleagues, starting in the late 1960s and extending
into the mid-1980s. See~\citet{frank2005network}, for example, for a
relatively recent survey of that literature. With the modern explosion
of interest in complex networks, there was a resurgence of interest in
sampling. Initially, the focus was on the simple awareness, and then
understanding of whether and how sampling affects the extent to which
the shape of the degree distribution of the observed network $G^*$
reflects that of the true network $G$. Seminal work during this period
includes an important empirical study by~\citet{lakhina2003sampling}, in
the context of traceroute sampling in the Internet, with follow-up
theoretical work by~\citet{achlioptas2005bias}, and work by Stumpf and
colleagues [e.g., \citet{stumpf2005sampling}, \citet{stumpf2005subnets}],
motivated, among other things, by networks arising in computational biology.

The focus on sampling of online social networks, as described above, is
arguably the most recent direction in this literature, with a flurry of
papers appearing in just the past five years. One of the first papers
to look closely at the implications of sampling in very large social
media networks (among others) was by~\citet{leskovec2006sampling}, where
attention was primarily on more classical network sampling designs
(e.g., so-called induced and incident subgraph sampling). This was
followed by papers like those by~\citet{hubler2008metropolis} and~\citet
{ribeiro2010estimating}, wherein samplers based on principles of the
Monte Carlo Markov chain were introduced and explored. Other examples
in this highly active area include \citet{ahn2007analysis}, \citet
{ahmed2010time}, \citet{ahmed2011network}, \citet{ahmed2012network}, \citet
{maiya2010online}, \citet{maiya2010sampling}, \citet{li2011sampling},
\citet{DBLP:journals/corr/abs-1109-1063}, \citet{shi2008very}, \citet
{mislove2007measurement}, \citet{lu2012sampling}, \citet{lim2011online},
\citet{gjoka2010walking}, \citet{gjoka2011multigraph}, \citet
{wang2011understanding}, \citet{zhou2011counting}, \citet
{kurant2011walking}, \citet{kurant2011towards}, \citet
{salehi2011characterizing}, \citet{mohaisen2012measuring}, and \citet
{jin2011albatross}.

In all of these papers, there is a keen interest in understanding the
extent to which characteristics of the network $G^*$ are reflective of
those of $G$. Typical characteristics of interest include degree
distribution, density, diameter, the distribution of the clustering
coefficient, the distribution of sizes of weakly (strongly) connected
components, Hop-plot, distribution of singular values (vectors) of the
network adjacency matrix, the graphlet distribution, the vertex (edge)
label density and the assortative mixing coefficient.

Here, in this paper, the network property we focus on is degree
distribution. The degree distribution of a network $G$, denoted by $\{
f_d\}$, specifies the proportion $f_d$ of vertices to have exactly $d$
incident edges, for $d=0,1,\ldots.$ It is arguably the most fundamental
quantity associated with a network and, importantly, one that may be
adversely affected by sampling, sometimes dramatically so [e.g.,
\citet{lakhina2003sampling,stumpf2005subnets}], hence, the following basic
question: how do we recover the degree distribution of some true
underlying network $G$, given only the information provided by the
sampled network $G^*$? For simplicity of exposition, hereafter we use
the term \emph{true degree distribution} and \emph{observed degree
distribution} to represent the degree distribution of $G$ and $G^{*}$,
respectively.

\citeauthor{frank1980estimation} (\citeyear{frank1980estimation}, \citeyear{frank1981survey}) shows that, under certain
network sampling designs, the expectation of the observed degree
relative frequencies is a linear combination of the true degree
relative frequencies. Let $\mathbf{f}=(f_k)$ and $\mathbf
{f}^*=(f_k^{*})$ be the vectors of true and observed degree frequencies
in $G$ and $G^{*}$, respectively. Then
%
\begin{equation}\label{eq:inv_freq}
E\bigl[\mathbf{f}^{*}\bigr] = \tilde{P} \mathbf{f},
\end{equation}
where $\tilde{P}$ depends fully on the sampling scheme and not on the
network itself. Thus, a natural unbiased estimator of $\mathbf{f}$
would seem to be simply $\tilde{P}^{-1}\mathbf{f}^*$. However, this
estimator suffers from two issues---$\tilde{P}$ typically is not
invertible in practice and, even when it is, $\tilde{P}^{-1}\mathbf
{f}^{*}$ may not be nonnegative.

From the perspective of nonparametric function and density estimation,
what we face is a linear inverse problem. One which, as we show, may
potentially be quite ill-posed, in the sense that the matrix $\tilde
{P}$ can be quite ill-conditioned. As a result, the estimation of
$\mathbf{f}$ must be handled with care, since naive inversion of
ill-conditioned operators in inverse problems typically will inflate
the ``noise'' accompanying the process of obtaining measurements, often
with devastating effects on our ability to recover the underlying
object (e.g., function or density). Here we offer, to the best of our
knowledge, the first principled estimator of a true degree distribution
$\mathbf{f}$ from a sampled degree distribution $\mathbf{f}^*$. In
particular, we propose a constrained, penalized weighted least squares
estimator, which, in particular, produces estimates that are
nonnegative (by constraint) and invert the matrix $\tilde{P}$ in a
stable fashion (by construction), in a manner that encourages smooth
solutions (through a penalty).

The rest of the paper is organized as follows. In Section~\ref{sec:char} we provide a detailed characterization of our inverse
problem, discussing the nature of the operator and the distribution of
noise. In Section~\ref{sec:solve} we describe our proposed approach to
solving this inverse problem, including a method for the automatic
selection of the penalization parameter. In Section~\ref{sec:simul} we
provide results of a simulation study, in which we study the impact on
the performance of our estimator of various parameters, including the
total number of vertices, the density of the network, sampling rates
and network types. In Section~\ref{sec:appl} we return to the primary
application of interest here, that of monitoring online social networks.
There we demonstrate the ability of our method to simultaneously
reconstruct accurately the degree distributions of various
sub-communities within online social networks corresponding to
Friendster, Orkut and LiveJournal. Finally, some additional discussion
and conclusions may be found in Section~\ref{sec:disc}.

\section{Characterizing the inverse problem}
\label{sec:char}

In solving inverse problems generally, it is important to understand
the nature of both the operator and the noise. Here the operator, in
the form of the matrix $\tilde{P}$, will derive entirely from the
network sampling design. At the same time, the ``noise'' (or, more
formally, the randomness in our measurements) also derives from the
sampling design. This linking of both operator and noise to our
sampling lends a certain element of uniqueness to our particular
inverse problem, the nature of which we aim to characterize in this section.

\subsection{Nature of the problem}
To begin with, assume we know the total number of vertices $n_v$ in the
underlying network. This is a reasonable assumption in the cases of,
for example, sampling a phone call network or surveying among a class
of students for their interactions. It is also not unreasonable in the
context of many online social networks where, for example, this may
either be readily available to those who own the network or reported to
the community as a basic summary statistic (e.g., the number of members
with active pages on Facebook). Thus, we know the degree distribution
$\mathbf{f}$ if and only if we know the degree counts $\mathbf
{N}=(N_0,N_1,\ldots,N_M)$, where $N_k$ is the number of vertices of
degree $k$, and $M$ is the maximum degree in the true network $G$. In
principle, the largest possible value for $M$ is $n_v-1$ in a simple
network where no multiple edges or self-loops exist, although in
practice we may have knowledge that it is smaller.

Under a given network sampling design, let $P(i,j)$ be the probability
that a vertex of degree $j$ in $G$ is selected and observed to have
degree $i$ in $G^*$. Following \citeauthor{frank1980estimation} (\citeyear{frank1980estimation,frank1981survey}), we will assume that the matrix
$P=[P(i,j)]$ of such probabilities depends only on the sampling design
and not, in particular, on the network $G$ itself. Then the equation
%
\begin{equation}
E\bigl[\mathbf{N}^*\bigr]=P\mathbf{N} \label{eq:inv_cnts}
\end{equation}
holds, in analogy to (\ref{eq:inv_freq}), where $\mathbf
{N}^*=(N_0^*,N_1^*,\ldots,N_M^*)$
is the vector of observed degree counts in $G^*$ and $P = \frac
{n_v^*}{n_v}\tilde{P}$ replaces $\tilde{P}$. Without loss of
generality, we will restrict our attention to this formulation of our
problem for the remainder of the paper.

It is useful to proceed with our characterization within the context of
the naive estimator of $\mathbf{N}$ obtained simply by inverting $P$,
that is,
%
\begin{equation}\label{eq:frank_est}
\hat{\mathbf{N}}_{\mathrm{naive}}=P^{-1}\mathbf{N}^*,
\end{equation}
where, again, we note that a formal inverse may or may not be
well-defined. The singular value decomposition (SVD) is a canonical
tool for studying the behavior of this estimator. Let $P=UDV^{T}$,
where $D=\operatorname{diag}(d_0,d_1,\ldots,d_M)$ is a diagonal matrix of
singular values, and $U=(\mathbf{u}_0,\mathbf{u}_1,\ldots, \mathbf
{u}_M)$, $V=(\mathbf{v}_0,\mathbf{v}_1,\break \ldots,\mathbf{v}_M)$ are
orthogonal matrices of the left- and right-singular vectors,
respectively. Then
%
\begin{equation}
\hat{\mathbf{N}}_{\mathrm{naive}} = \sum_{i=0}^M
\biggl[ \frac
{1}{d_i}\mathbf{u}_i^T \mathbf{N}^*
\biggr] \mathbf{v}_i \label{eq:frank_decomp}
\end{equation}
decomposes the naive estimator (\ref{eq:frank_est}) into a linear
combination of the right singular vectors of $P$.
\begin{figure}

\includegraphics{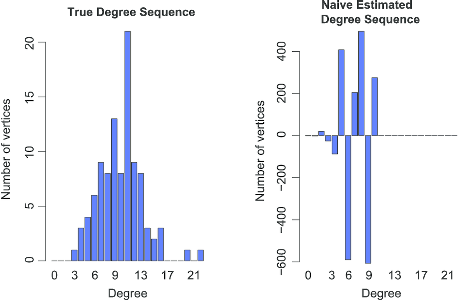}

\caption{Left: ER graph with 100 vertices and 500 edges. Right: Naive
estimate of degree distribution, according to equation (\protect\ref{eq:frank_est}).
Data drawn according to induced subgraph sampling with sampling rate
$p=60\%$.}
\label{fig:pure_frank}
\end{figure}

The quality of this estimator is determined, in part, by the extent to
which the vector $\mathbf{N}$ may be approximated well by such linear
combinations. In general, the right singular vectors $v_i$ vary in
smoothness, from smoother behavior (i.e., low-frequency) at small
values of $i$ to less smooth behavior (i.e., high-frequency) at larger
values of $i$. Since most degree distributions encountered in practice,
as well those induced through common choices of random graph models
(some examples of which we use in Section~\ref{sec:simul}), are
relatively smooth, typically with either exponential or power-law
behavior in the tails, intuitively it is the first handful of right
singular vectors upon which a sensible estimator should be based.
The stability of this estimator can be summarized through the condition
number of $P$, that is, the ratio of the largest to smallest singular
values. Larger condition numbers suggest greater instability in the
estimator. Intuitively, for unstable matrices $P$, the singular values
$d_i$ at higher indices $i$ are, comparatively, quite small. As a
result, the estimator in (\ref{eq:frank_decomp}) will put
disproportionately large weight on contributions from the latter (i.e.,
high-frequency) singular vectors. The end result is an estimator that
can oscillate in a decidedly unappealing manner, as illustrated in
Figure~\ref{fig:pure_frank}.

Since the operator $P$ plays such an important role in both the shape
and the stability of the estimator (and, by extension, more sensible
modifications of the estimator, such as we offer below), and $P$ in
turns is determined by the sampling design, we examine a handful of
canonical examples of sampling designs and their operators in the
following subsection.

\subsection{Common network sampling designs and the operator $P$}

Here we look at a few common network sampling designs and their
corresponding $P$ matrix. We consider them ordered from simpler to more
complex. We refer readers to
\citeauthor{kolaczyk2009statistical} (\citeyear{kolaczyk2009statistical},
Chapter~5)
for additional background on network sampling and a more comprehensive
list of sampling designs.

\subsubsection{Ego-centric and one-wave snowball sampling} Ego-centric
sampling (also called unlabeled star sampling) is a simple, nonadaptive
(conventional) sampling design. As \citet{handcock2010modeling} write
that ``[a] sampling design is conventional if it does not use
information collected during the survey to direct subsequent sampling
of individuals$\ldots$ [and] a sampling design [is] adaptive if it uses
information collected during the survey to direct subsequent sampling,
but the sampling design depends only on the observed data.'' Under
ego-centric sampling, first a set of vertices is selected according to
independent $\operatorname{Bernoulli}(p)$ trials at each vertex. Then all edges
incident to the selected vertices are observed. In this case, the
operator $P$ is a diagonal matrix with the sampling rate $p$ at each
diagonal position, that is,
%
\begin{equation}\label{eq:p_ego}
P_{\mathrm{ego}}(i,j) =
\cases{ p,  & $\quad\mbox{for } i=j=0,1,\ldots,M$,\vspace*{3pt}\cr
0,  &\quad $\mbox{for } i,j=0,\ldots, M;  i \neq j$.}
\end{equation}

A natural extension of this concept is one-wave snowball sampling.
Here, after an initial selection of vertices, there is a subsequent
selection of additional vertices, using the information obtained from
the initial selection. Therefore, one-wave snowball sampling is an
adaptive sampling design. The initial selection is again done according
to independent $\operatorname{Bernoulli}(p)$ trials. The subsequent selection contains
all vertices that have at least one connection with a vertex in the
initial set. Similar to ego-centric sampling, all edges incident to
vertices selected in either of the two sets are then observed, so the
operator $P$ is again a diagonal matrix, with entries
%
\begin{equation}\label{eq:p_snow}
P_{\mathrm{snow}}(i,j) =
\cases{ 1-(1-p)^{i+1},  &\quad$\mbox{for } i=j=0,1,\ldots, M$,\vspace*{3pt}
\cr
0,  & \quad$\mbox{for } i,j=0,\ldots, M; i\neq j $.}
\end{equation}

These two sampling designs (as well as multi-wave snowball sampling and
other variations) are common in social network studies, where, for
example, a selection of individuals are interviewed and asked to
nominate their connections or partners. Readers can refer to~\citet{rolls2012modelling} for more details, in the context of networks of
injecting drug users. We note that the adaptive designs we consider
here are the textbook versions and not complicated adaptations that
might sometimes be used in practice due to resource limitations for
following links. Even so, the standard and simple designs we consider
with known and constant matrix $P$ would be the logical point of
departure for research on correcting the sampling bias of the degree
distribution in more complex adaptive designs.

For a diagonal $P$ matrix, the singular values are equal to the
diagonal elements. Both the left and right singular vectors are the
canonical set of basis vectors $\{e_i\}_{i=1}^{M+1}$, where $e_i$
contains a $1$ at the $i$th entry and $0$ at all the other entries.
Since $P_{\mathrm{ego}}=I\times p$, where $I$ is the identity matrix,
$P_{\mathrm{ego}}$ is not ill-conditioned at all. To estimate the degree
count vector $\mathbf{N}$, we need only scale the observed degree count
vector $\mathbf{N}^*$ by $1/p$. That is, the naive estimator is $\hat
{N}_{\mathrm{naive}}=N^*/p$.

In one-wave snowball sampling, the observed degree counts are biased,
because in the second round of vertex selection, there is more chance
to select the vertices that have more connections. The observed degree
count vector therefore can be thought of as moving to the right of the
true degree count vector. Hence, at a minimum, a good estimator should
correct the observations by moving the distribution back to the left.
How difficult this task may be is summarized by the condition number of
$P_{\mathrm{snow}}$, which is equal to
%
\begin{equation}\label{eq:cond.num.snow}
\frac{P_{\mathrm{snow}}(M,M)}{P_{\mathrm{snow}}(0,0)}=\frac{1-(1-p)^{M+1}}{
1-(1-p)}= \frac{1-(1-p)^{M+1}}{p},
\end{equation}
and therefore depends on the relationship between the expected
proportion $p$ of vertices sampled initially and the maximum degree
$M$. In the case where $p$ is fixed, as $M$ increases, the condition
number is upper bounded by $\frac{1}{p}$. On the other hand, if $Mp =
o(1)$, using the approximation
$(1-p)^{M+1} \approx 1 - (M+1)p$, we find that the condition number
behaves as $(M+1)$.

These observations suggest that, for instance, under low sampling rates
the inverse problem is increasingly ill-posed for estimating degree
distributions of heavier tails. Also, the bounds on the condition
numbers suggest that, in contrast to estimation of the mean from a
sample from a finite population, where the accuracy depends on the
sample size rather than the fraction of the population that is sampled,
for estimation of complex properties of networks the accuracy depends
strongly on the fraction of the population that is sampled.


\subsubsection{Induced and incident subgraph sampling} These two
sampling designs are both nonadaptive and analogous in spirit,
differing only in the order of selection of vertices and edges. In
induced subgraph sampling, a set of vertices is selected as independent
$\operatorname{Bernoulli}(p)$ trials (other variations are possible---see below).
Then, all edges between selected vertices are observed, that is, we
observe the subgraph induced by this vertex subset. This sampling
scheme has been used in the analysis of technological and biological
networks [\citet{stumpf2005sampling}]. Conversely, under incident
subgraph sampling we select \textit{edges} as independent $\operatorname{Bernoulli}(p)$
trials and we then observe all vertices incident to at least one
selected edge.

The $P$ matrix for induced subgraph sampling is
%
\begin{equation}\label{eq:p_ind}
P_{\mathrm{ind}}(i,j) =
\cases{ \ds \pmatrix{j \cr i}
p^{i+1}(1-p)^{j-i},  &\quad$\mbox{for } 0\le i \le j \le M$,
\vspace*{3pt}\cr
0,  & \quad$\mbox{for } 0\le j < i \le M$,}
\end{equation}
while that for incident subgraph sampling is
%
\begin{equation}\label{eq:p_inc}
P_{\mathrm{inc}}(i,j) =
\cases{ \ds \pmatrix{j \cr i}
p^{i}(1-p)^{j-i},  &\quad$\mbox{for } 1\le i \le j \le M$,\vspace*{3pt}
\cr
0,  & \quad$\mbox{for } 0\le j < i \le M$.}
\end{equation}

Notice that for incident subgraph sampling the index $i$ starts from
$1$, because there are no isolated vertices in the sample.

These two sampling designs are widely studied in literature, for
example, in \citet{stumpf2005sampling}, \citet{leskovec2006sampling},
\citet{ahmed2011network}, and \citet{kurant2012coarse}, to name a few. In
some cases, simple random sampling (SRS) is used instead of Bernoulli
sampling to select the initial vertices or edges. However, under
appropriate calibration of $p$, the former can be well approximated by
the latter for large networks and small to moderate $p$. So, without
loss of generality, we ignore this variant for the purposes of exposition.

\begin{figure}[b]

\includegraphics{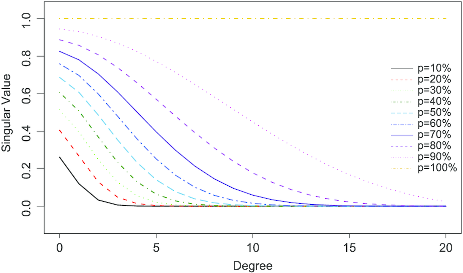}

\caption{Singular values decay under induced subgraph sampling. $M=20$.}
\label{fig:singular_values.ind}
\end{figure}

Unlike ego-centric and one-wave snowball sampling, the structure of the
operator under induced/incident subgraph sampling can cause severe
problems if we try to invert it naively. Because the structure of
$P_{\mathrm{inc}}$ is very similar to $P_{\mathrm{ind}}$, we only analyze
$P_{\mathrm{ind}}$ here. The condition number in this case is equal to
$p^{-M}$ and so, as the sampling rate $p$ goes down or the maximum
degree $M$ increases, the operator $P$ becomes more ill-conditioned. In
real-world situations, such as the monitoring of online social
networks, sampling rates are typically low (e.g., 10--20\%) and $M$ is
typically large (e.g., on the order of 100's or 1000's), and thus $P$
is decidedly ill-conditioned and effectively not invertible. The
overall pattern of decay of the singular values under induced subgraph
sampling is illustrated
in Figure~\ref{fig:singular_values.ind}.

\begin{figure}

\includegraphics{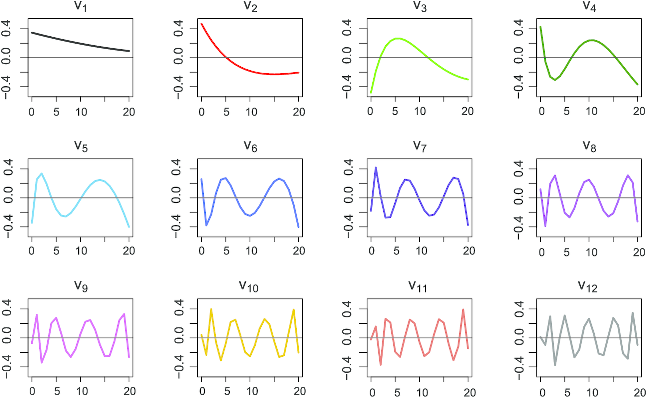}

\caption{The first 12 right singular vectors under induced subgraph
sampling, ordered by singular values from big to small: maximum degree
$M=20$, sampling rate $p=20\%$.}
\label{fig:V}
\end{figure}

\begin{figure}[b]

\includegraphics{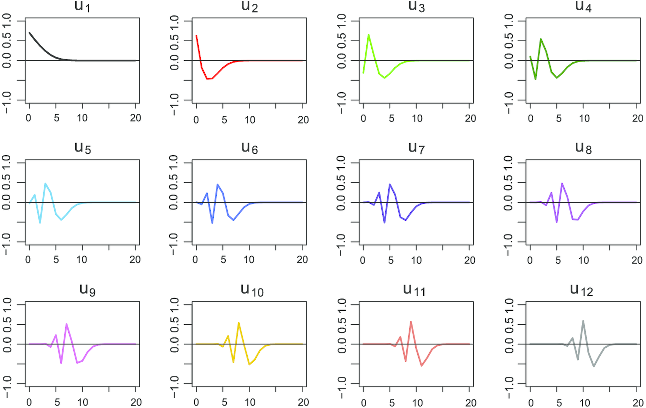}

\caption{The first 12 left singular vectors under induced subgraph
sampling, ordered by singular values from big to small: maximum degree
$M=20$, sampling rate $p=20\%$.}
\label{fig:U}
\end{figure}

Recall that the decomposition in (\ref{eq:frank_decomp}) shows the
naive estimator to be a linear combination of the right singular
vectors $v_i$, with weights determined in part by the inner product of
the observations $\mathbf{N}^*$ with the left singular vectors $u_i$.
Examination of these vectors can provide additional insight into the
expected behavior of this estimator. As can be seen from the
illustration in Figure~\ref{fig:V}, the right singular vectors behave
like a Fourier basis, in that they are supported over the full range of
degrees $k$ and oscillate increasingly with higher indices $i$. On the
other hand, the left singular vectors, shown in Figure~\ref{fig:U},
behave in a more stable fashion with increasing index $i$, with only
the support changing noticeably at the higher indices, moving like a
window from low degrees $k$ to high. Combined with our previous
observation of the drastic decay in singular values $d_i$, this
explains the behavior of the estimate in Figure~\ref{fig:pure_frank}.

While it would be desirable to have an analytical expression for the
singular vectors under induced subgraph sampling, we are unable to
produce one; however, it is possible to produce expressions for the
eigenfunctions of $P_{\mathrm{ind}}$, as solutions to the nonsymmetric
eigen-decomposition $P_{\mathrm{ind}} = \tilde{U} \Lambda\tilde{U}^{-1}$. These
do not appear to be helpful in yielding similarly interpretable
expressions for the SVD but, nonetheless, may be of some independent
interest. We therefore include this result in Appendix~\ref{appen:eigen}.


\subsubsection{Random walk and other exploration-based methods}
Another
class of sampling plans that has arisen recently, and has been of
particular interest to the community working with online social
networks, is that based on notions of visiting vertices and edges in a
network in the course of a random walk on the graph~$G$. Specifically,
in the basic version of random walk sampling, we first select a vertex~$u$ uniformly at random from $V$. Then one of $u$'s neighbor vertices,
say $v$, is chosen uniformly at random from the set of $u$'s neighbors.
In turn, one of $v$'s neighbor vertices, say $w$, is chosen uniformly
at random from the set of $v$'s neighbors. The process is repeated, and
the selected vertices $\{u,v,w,\ldots\}$ along with the edges $\{(u,v),(v,w),\ldots\}$ constitute the sample. For examples of other
members of this family, we refer readers to~\citet{leskovec2006sampling}
and~\citet{ribeiro2010estimating}.

If we consider a random walk sampling over a nonbipartite, connected,
undirected graph, once the steady state is reached, it shares an
important property with incident subgraph sampling with SRS of edges,
in that both sample edges uniformly at random [\citet{ribeiro2010estimating}]. Thus,
%
\begin{equation}\label{eq:p_rw}
P_{\mathrm{RW}}(i,j)=
\cases{ \ds \pmatrix{j \cr i} \pmatrix{n_e-j
\cr n^*_e-i} \pmatrix{n_e \cr n^*_e}^{-1},
&\quad$\mbox{for } 1\le i \le j \le M$,
\vspace*{3pt}\cr
0,  &\quad$\mbox{for } 0\le j < i \le M$,}
\end{equation}
where $n_e$ is the total number of edges in the true network and
$n^*_e$ is the number of edges selected in the sample. Therefore, with
respect to the nature of the inverse problem that we study here, we may
categorize this sampling plan with the induced and incident subgraph
sampling plans described above.


\subsection{Distribution of the noise}
\label{sec:noise}

The observation $\mathbf{N}^*$ can be viewed as a ``noisy'' version of
$N$. However, as remarked earlier, since it is assumed here that there
is no measurement error (e.g., if a query of Facebook indicates person
$A$ has ``friended'' person $B$, then we accept that they have), the
``noise'' is rather a reflection of the randomness due to sampling.
Because we intend to pursue a regression-based approach to solving our
linear inverse problem, the question of what noise model to use as an
approximation to sampling variability is important. We discuss this
question now.

For ego-centric sampling, a vertex is observed to have degree $k$ if
and only if the vertex is selected through Bernoulli sampling and also
has degree $k$ in the true graph. Therefore,
%
\begin{equation}\label{eq:pois_appro.ego_snow}
N_k^* = \sum_{\{u\dvtx  d_u=k\}} I\bigl\{u\in
V^{*}\bigr\},
\end{equation}
where $d_u$ represents the degree of a vertex $u \in V$ in $G$, and
$d_u^*$ represents the degree of a vertex $u \in V^{*}$ in $G^{*}$. For
each $k$, there are $N_k$ such independent indicator functions, and
each indicator function has the same probability to be one. Thus, the
distribution of the $N_k^*$ is that of $M+1$ independent binomials,
that is, $N_k^* \sim\operatorname{Bin} (p,N_k)$. For small $p$ and large
$N_k$, we can expect that these binomials may be well-approximated as
Poisson random variables, with means $N_k   p$.

The case of one-wave snowball sampling and induced subgraph sampling
(as well as the related cases of incident subgraph sampling and random
walk sampling) is decidedly less straightforward to analyze. The
expectation of $\mathbf{N}^*$ is, of course, provided by equation
(\ref{eq:inv_cnts}). The variance (covariance) formula is more complicated.

For one-wave snowball sampling, the representation (\ref{eq:pois_appro.ego_snow}) still applies. However, the indicator
functions are not independent. Straightforward arguments yield that the
covariance and variance of $N_k^*$ for $k = 0,1,\ldots,M$ are
%
\begin{eqnarray}\label{eq:cov.snow}
%
\operatorname{Cov}\bigl(N_k^*,N_l^*\bigr)&=&\sum_t N_{1klt}
\bigl[1-(1-p)^{l+1}-(1-p)^{k+1}+(1-p)^{k+l-t}\bigr]
 \nonumber\\
&&{}+\sum_t N_{0klt}\bigl[1-(1-p)^{l+1}-(1-p)^{k+1}+(1-p)^{k+l-t+2}\bigr]\\
&&{}-N_kN_l P_{\mathrm{snow}}(k,k)P_{\mathrm{snow}}(l,l) \nonumber
\end{eqnarray}
and
%
\begin{eqnarray}\label{eq:var.snow}
%
\operatorname{Var}\bigl(N^*_k\bigr) &=& N_kP_{\mathrm{snow}}(k,k) \nonumber \\
&&{}+\sum_t N_{1kkt} \bigl[1-2(1-p)^{k+1}+(1-p)^{2k-t}\bigr]
\nonumber
\\[-8pt]
\\[-8pt]
\nonumber
&&{}+\sum_t N_{0kkt} \bigl[1-2(1-p)^{k+1}+(1-p)^{2k-t+2}\bigr] \\
&&{}-\bigl(N_k P_{\mathrm{snow}}(k,k) \bigr)^2, \nonumber
\end{eqnarray}
where $N_{0klt}$ ($N_{1klt}$) is determined by the underlying network
$G$, defined as the number of ordered pairs of nonadjacent (adjacent)\vadjust{\goodbreak}
distinct vertices of degrees $k$ and $l$, respectively, which have $t$
common adjacent vertices.

For induced-subgraph sampling, we can write
%
\begin{equation}\label{eq:pois_appro.ind}
N_k^* = \sum_{r=k}^{M}\sum
_{u=1}^{n_v} I\bigl\{u\in V^{*},
d_u^*=k, d_u=r\bigr\}.
\end{equation}
Using arguments analogous to those in \citet{frank1980estimation}, it is
possible to show that, for $k=0,1,\ldots, M$, the
variance takes the form
\begin{eqnarray}\label{eq:var.ind}
%
\operatorname{Var}\bigl(N^*_k\bigr) &=& \sum_iN_i P_{\mathrm{ind}}(k,i) \nonumber \\                         
&&{}+\sum_r\sum_s\sum_t N_{0rst}\sum_m\pmatrix{t \cr m }
\pmatrix{r-t \cr k-m} \pmatrix{s-t \cr k-m}\nonumber\\
&&\hspace*{100pt}{}\times p^{2k-m+2}q^{(r+s-t)-(2k-m)}
\nonumber
\\[-8pt]
\\[-8pt]
\nonumber
&&{}+\sum_r\sum_s\sum_t N_{1rst}\sum_m\pmatrix{t \cr m } \pmatrix{r-t-1 \cr k-m-1} \pmatrix{s-t-1 \cr k-m-1}\\
&&\hspace*{100pt}{}\times  p^{2k-m}q^{(r+s-t)-(2k-m)} \nonumber \\
&&{}-(\sum_i  N_i P_{\mathrm{ind}}(k,i))^2 .  \nonumber
\end{eqnarray}
Using similar techniques, it is also possible to write out a similar
formula for $\operatorname{Cov} (N_j^*,N_k^* )$, which we find is, in
general, nonzero for $j \neq k$, as would be expected.

Now consider the marginal distributions of the $N^*_k$ under snowball
sampling and induced subgraph sampling. Note that the first term in
(\ref{eq:var.snow}) and (\ref{eq:var.ind}) is the $k$th entry of the
expectation $P\mathbf{N}$. This observation suggests that, if the
remaining terms in the variance (as well as the off-diagonal terms
corresponding to covariances) are sufficiently small, a Poisson model
might again be acceptable.

More precisely, if the sampling rate $p$ is small, then each of the
indicators in~(\ref{eq:pois_appro.ego_snow}) and (\ref{eq:pois_appro.ind}) likely has only very small probability of being
equal to one. On the other hand, if the graph is large (i.e., $n_v$ is
large) and $k$ is not too far out in the tail of the distribution
(i.e., $k$ is not too close to $M$), then there should be many such
indicators. So a Poisson approximation would make sense here. Given,
however, that these indicator variables are dependent, the necessary
argument is somewhat more involved. We present a formal justification,
using the Chen--Stein method, in Appendix~\ref{appen:pois}.

\begin{figure}

\includegraphics{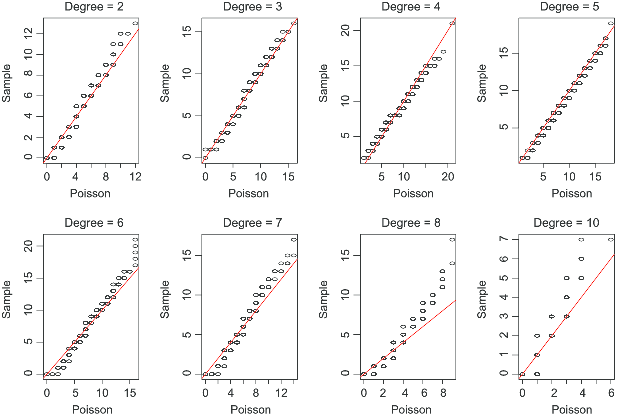}

\caption{QQ plot: distribution of $N_i^*$ compared to Poisson
distribution with mean $(PN)_i$. The underlying network is ER with
$n_v=|V|=1000$ and $n_e=|E|=50{,}000$. Sampling rate $p=5\%$. The average
degree of the sample is equal to $5$.}
\label{fig:qq_pois}
\end{figure}

Simulation can be used to assess this approximation. Some
representative results, shown in Figure~\ref{fig:qq_pois}, confirm the
reasonableness of a Poisson approximation for the marginal distribution
of the $N^*_k$, under induced subgraph sampling, for $k$ within a
reasonable distance from the mean.
%

\begin{figure}

\includegraphics{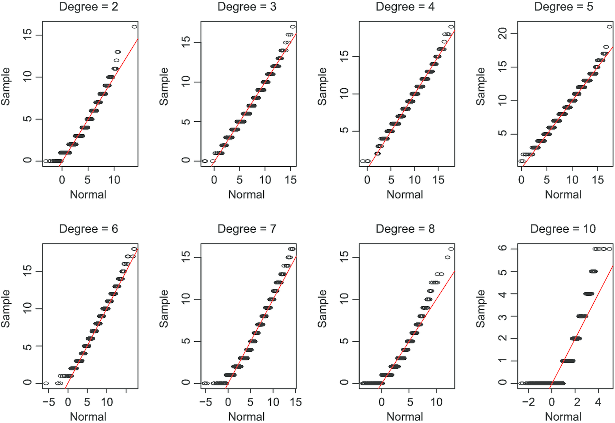}

\caption{QQ plot: distribution of $N_i^*$ compared to Gaussian
distribution with mean $(PN)_i$ and sample variance. The underlying
network is ER with $n_v=|V|=1000$ and $n_e=|E|=50{,}000$. Sampling rate
$p=5\%$. The average degree of the sample is equal to $5$.}
\label{fig:qq_Gaussian}
\end{figure}

In summary, for all of the sampling plans considered in this paper, an
approximate Poisson marginal distribution is arguably reasonable for
the observed counts~$N^*_k$. Thus, a Poisson regression model is
suggested for solving our inverse problem. However, for reasons of
numerical efficiency and stability, we prefer to approximate this model
in turn by a Gaussian model, with nonconstant variance that varies in
proportion to the mean, leading to a weighted least squares regression.
Simulation results (shown in Figure~\ref{fig:qq_Gaussian}) suggest that
this, too, is a reasonable choice. Accordingly, our model development,
as described starting in the next section, will implicitly assume a
Gaussian noise model. 

\subsection{Discussion of assumptions}\label{sec24}

In some sampling designs, nodes' inclusion probabilities can depend on
unobserved properties of the node, such as its true degree, or on other
unobserved properties of the network. In this paper we restrict
attention to sampling designs (ego-centric, one-wave snowball sampling,
induced/incident subgraph sampling, random walk) where inclusion
probabilities are known. This restriction underlies (\ref{eq:inv_freq})
and (\ref{eq:inv_cnts}) to be established without the need for
assumptions about the structure of the network itself. The approach we
take is called ``design-based'' in the sampling literature, as compared
to ``model-based.'' \citet{handcock2010modeling} observe the following:

\begin{quote}
In the \emph{design-based} framework [$G$] represents the
fixed population and interest focuses on characterizing based on
partial observation. The random variation considered is due to the
sampling design alone. A key advantage of this approach is that it does
not require a model for the data themselves$\ldots$ Under the \emph{model-based} framework, $[G]$ is stochastic and is a realization from
a stochastic process depending on a parameter $\eta$. Here interest
focuses on $\eta$ which characterizes the mechanism that produced the
complete network~$[G]$.
\end{quote}

Design-based inferences are generally not feasible (i) for adaptive
sampling designs other than a network census and ego-centric sampling
designs [\citet{handcock2010modeling}, 11ff] or (ii) for any designs for
which the inclusion probabilities of sampled nodes (and dyads, triads,
etc., depending on the application) are unknown at least up to a
scaling factor. Design-based inference is the standard mode for
analysis of samples obtained by government statistical agencies or for
large-scale random samples funded by government agencies. That is not
to say that assumptions are not brought in for taking into account
nonresponse or response error, but the latter two sources of error
depend on the properties of the sampled units rather than the sampling
design itself. Although design-based inference is applicable only to a
restricted set of sample designs, it has the advantage of not requiring
specific knowledge about the graph or network being sampled.

We are assuming that the number of nodes is known, consistent with the
only other research on design-based inferences for the degree
distribution. The assumption is not strictly necessary, as the number
of nodes is estimable by a Horvitz--Thompson estimator for the designs
under consideration [\citet{handcock2010modeling}, pages 12--13], but the
assumption simplifies the exposition. We also assume that the sampling
probabilities of nodes (or edges) are known, which is a standard
assumption for conventional sampling designs~[e.g., \citet
{cochran1977g}] and not unrealistic for the designs we are considering.

We assume as well that the nodes and edges in the sample are observed
without error. In the network literature, the question of effect of
such observational error and how to quantify and adjust for it is still
largely unexplored, and hence is beyond the scope of this paper.

\section{Estimating the degree distribution}
\label{sec:solve}


Bearing in mind the SVD-based representation of the naive estimator
$P^{-1}\mathbf{N}^*$ of $\mathbf{N}$, as shown in (\ref{eq:frank_decomp}), the analyses of Section~\ref{sec:char} together
suggest that a better solution to our inverse problem would be an
estimator developed in a manner analogous to ridge regression and other
similar penalized regression strategies. In this section, we offer such
an approach.

We adopt a penalized least squares perspective in defining our
estimator. Informed by our analysis of the ``noise'' in our inverse
problem, we specify a generalized least squares criterion. Furthermore,
since the vector of degree counts should be everywhere nonnegative
and, additionally, the total degree counts should equal the total
number of vertices, $n_v$, we include these two properties as
constraints. Our estimator $\hat{\mathbf{N}}$ for $\mathbf{N}$ is then
the solution to the following optimization problem:
%
\begin{eqnarray}
&& \mathop{\operatorname{minimize}}_{\mathbf{N}} \qquad   \bigl(P
\mathbf{N}-\mathbf{N}^* \bigr)^TC^{-1} \bigl(P\mathbf {N}-
\mathbf{N}^* \bigr)+\lambda\cdot \operatorname{pen}(\mathbf{N})\nonumber
\\
\label{eq:opt}
&& \mbox{subject to}\qquad  N_i \geq0, i=0,1,\ldots M,
\\
\nonumber
&&\phantom{\mbox{subject to}}\qquad \sum_{i=0}^M{N_i}=n_v,
\end{eqnarray}
where $C$ denotes the covariance matrix of $\mathbf{N}^*$, that is,
$C=\operatorname{Cov}(\mathbf{N}^*)$, $\operatorname{pen}(\mathbf{N})$ is a penalty on
the complexity of $\mathbf{N}$, and $\lambda$ is a smoothing parameter.

Under a convex penalty, (\ref{eq:opt}) has the canonical form of a
convex optimization [\citet{boyd2004convex}] and, in principle, standard
software can be used. For example, CVX, a package for specifying and
solving convex programs [\citet{cvx}], can be used to solve
(\ref{eq:opt}). In our case, because we use a penalty based on an $\ell_2$
norm, as discussed below, (\ref{eq:opt}) can be written as a quadratic
programming problem. Accordingly, we use \textit{quadprog}, the quadratic
programming function in the MATLAB optimization toolbox, to solve (\ref{eq:opt}).

Note that the solution spaces of the original problem (\ref{eq:inv_cnts}) and (\ref{eq:opt}) are not the same.
The solution (\ref{eq:frank_est}) of the original problem (\ref{eq:inv_cnts}) is a point
in a space generated by the right singular vectors $\{v_i\}$. The
constraint\vspace*{1.5pt} and penalized solution of~(\ref{eq:opt}) is a point in a space
generated by $\{B^{-1}v_i\}$, where $B=
\bigl[{\fontsize{8.36}{9.36}\selectfont{\matrix{P^TC^{-1}P+\lambda\Omega & \tfrac{1}{2}\mathbf{1}\vspace*{-0.5pt}\cr
\mathbf{1}^T &  0}}}\bigr]$, ignoring the nonnegativity constraint as is shown in~(\ref{C6}). Through
this we obtain smoothing.

In the following subsections we discuss choice of the penalty,
selection of the smoothing parameter and various practical considerations.

\subsection{Penalty}

There are a variety of penalties common in the literature on
nonparametric function estimation, usually consisting of a norm (e.g.,
$\ell_1$, $\ell_2$, total-variation, etc.) applied to some functional
of the proposed estimator. The choice of penalty should reflect the
assumption of smoothness, that is, $f_k \approx f_l$ if $k$ and $l$ are
close. Examples of networks with smooth degree distributions include
Erd\"{o}s-R\'{e}nyi (ER), mixture of ER, power-law networks, networks
having exponential or power-law tails, as well as those having the body
of the exponential or power-law networks. We want to force our
estimates toward distributions with such smoothness, where the naive
estimates have obvious flaws (e.g., Figure~\ref{fig:pure_frank}).

In our framework,\vspace*{1pt} the assumption of a smooth true degree distribution
is accounted for by choosing a penalization of the form $\|D\mathbf{N}\|
_2^2$, where the matrix $D$ represents a second-order differencing
operator. Specifically, the formula for $D$ is
%
\begin{equation}
D =
\left[
\matrix{ 1 & -2 & 1 & 0 & \cdots& 0 & 0 & 0 & 0
\vspace*{2pt}\cr
0 & 1 & -2 & 1 & \cdots& 0 & 0 & 0 & 0
\vspace*{2pt}\cr
0 & 0 & 1 & -2 & \cdots& 0 & 0 & 0 & 0
\vspace*{2pt}\cr
\vdots& \vdots& \vdots& \vdots& & \vdots& \vdots& \vdots& \vdots
\vspace*{2pt}\cr
0 & 0 & 0 & 0 & \cdots& -2 & 1 & 0 & 0
\vspace*{2pt}\cr
0 & 0 & 0 & 0 & \cdots& 1 & -2 & 1 & 0
\vspace*{2pt}\cr
0 & 0 & 0 & 0 & \cdots& 0 & 1 & -2 & 1 } \right].
\end{equation}

This choice, in the discrete setting, is analogous to the use of a
Sobolev norm with nonparametric function estimation in the continuous
setting. It assumes mean-square curvature of the degree distribution is
small. This is one commonly used smoothing regularization, and we have
found it to work well with the types of degree distributions explored
here. Other penalties may work less well. For example, the $L1$ norm
can be used as a heuristic for finding a sparse solution, thus the
solutions $\hat{N}$ can be truncated. We refer readers to Chapter~6.6.6
of \citet{boyd2004convex} for how different penalty functions perform
generally on denoising problems.

\subsection{Selection of the penalization parameter \texorpdfstring{$\lambda$}{lambda}}
Denote the solution to the optimization problem in (\ref{eq:opt}) as
$\hat{\mathbf{N}}=f_{\lambda}(\mathbf{N}^*)$, a function of $\mathbf
{N}^*$, indexed by $\lambda$. For a given observation vector $\mathbf
{N}^*$, a bigger $\lambda$ produces a smoother estimator. The problem
of selecting an optimal $\lambda$ falls into the category of model
selection. However, commonly used cross-validation methods which assume
independent and identically distributed observations do not apply to
our network sampling situation because, as already discussed, the
$N_i^*$ for $i=0,\ldots,M$ are not identically distributed and there
are nonzero correlations between $N_i^*$ and $N_j^*$ for $i\neq j$.
Instead, we offer a strategy based on the method of generalized Stein's
unbiased risk estimation (SURE), proposed in \citet{eldar2009generalized}.

We define a weighted mean square error (WMSE) in the observation space as
%
\begin{equation}\label{eq:mse}
\mathrm{WMSE}(\hat{\mathbf{N}}, \mathbf{N})=E \bigl[ (P\mathbf{N}-
P\hat{\mathbf{N}})^T C^{-1}(P\mathbf{N}-P\hat{\mathbf{N}}) \bigr].
\end{equation}
Under the conditions that $f_{\lambda}(\mathbf{N}^*)$ is weakly
differentiable and that $E\llvert  f_{\lambda}(\mathbf{N}^*)\rrvert $ is
bounded (which we verify following the arguments in Appendix~\ref{appen:sure}), a generalized SURE estimate for the WMSE can be obtained as
%
\begin{eqnarray}
\widehat{\mathrm{WMSE}}(\hat{\mathbf{N}},\mathbf{N})&=& (P\mathbf{N})^TC^{-1}P
\mathbf{N}+(P\hat{\mathbf{N}})^TC^{-1}P\hat{\mathbf{N}}
\nonumber
\\
\label{eq:WMSE}
&&{}+2 \biggl\{\operatorname{Trace} \biggl(P\frac{\partial\hat{\mathbf{N}}}{\partial\mathbf{N}^*} \biggr) \biggr\}
\\
&&{}-2(P\hat{\mathbf{N}})^TC^{-1}\mathbf{N}^*.
\nonumber
\end{eqnarray}

The first term in (\ref{eq:WMSE}) involves the unknown $\mathbf{N}$.
However, we may drop this term because it does not involve $\lambda$.
The last three terms have $\hat{\mathbf{N}}$ in them, which is a
function of $\lambda$. Given $P, \mathbf{N}^*$ and $C$ as well, the
second and fourth terms are straightforward to compute.
The third term, called the divergence term in \citet{eldar2009generalized}, can be simulated using the Monte Carlo
technique proposed in \citet{ramani2008monte}. Specifically, let $b$ be
a vector with zero mean, covariance matrix $I$ (i.e., independent of
$\mathbf{N}^*)$ and bounded higher order moments. Then
%
\begin{equation}
\operatorname{div}\equiv\operatorname{Trace} \biggl(P\frac{\partial\hat{\mathbf{N}}}{\partial\mathbf{N}^*} \biggr)=\lim
_{\varepsilon\to0} E_{b} \biggl\{ b^TP \biggl(
\frac{f_\lambda (\mathbf{N}^*+\varepsilon\mathbf{b}
)-f_\lambda (\mathbf{N}^* )}{\varepsilon} \biggr) \biggr\}.
\end{equation}

Let $\mathbf{b_i}$ be the realization of $\mathbf{b}$ at each
simulation. The algorithm for estimating $\operatorname{div}=\operatorname{Trace} (P\frac
{\partial\hat{\mathbf{N}}}{\partial\mathbf{N}^*} )$ and
computing of $\widehat{\mathrm{WMSE}}$ for a given $\lambda= \lambda_0$ and
fixed $\varepsilon$ is as follows:
\begin{enumerate}
\item $\mathbf{y}=\mathbf{N}^*$;
\item For $\lambda= \lambda_0$, evaluate $f_{\lambda}(\mathbf{y})$; $i
= 1$; $\operatorname{div} = 0$;
\item Build $\mathbf{z} = \mathbf{y} + \mathbf{b_i}$; evaluate\vspace*{1pt}
$f_{\lambda}(\mathbf{z})$ for $\lambda= \lambda_0$;
\item $\operatorname{div}=\operatorname{div}+\frac{1}{\varepsilon} \mathbf{b_i}^TP(f_{\lambda} (\mathbf
{z})-f_{\lambda} (\mathbf{y}));i=i+1$;
\item If $(i \leq K)$\vspace*{1pt} go to Step 3; otherwise evaluate sample mean:
$\operatorname{div}= \operatorname{div}/K$ and compute $\widehat{\mathrm{WMSE}}(\lambda_0)$ using (\ref{eq:WMSE}).
\end{enumerate}
We offer recommendations for the practical selection of $\varepsilon$ and
$K$, as well as the distribution of $b$, in Section~\ref{sec:simul}.

For\vspace*{1pt} a fixed $\mathbf{N}^*$, by minimizing $\widehat{\mathrm{WMSE}}$ with
respect to $\lambda$, we find the optimal $\lambda$ that minimizes
$\widehat{\mathrm{WMSE}}$.

\subsection{Approximation of the covariance matrix $C$}\label{sec33}

For the ego-centric sampling design, recall that the $N^*_k$ are
independent random variables, distributed according to a binomial with
parameters $p$ and $N_k$. As a result, the covariance matrix $C$ is
simply $p (1-p)\times\operatorname{diag}(\mathbf{N})$. In contrast, for the
one-wave snowball sampling and the induced subgraph sampling (as well
as the related incident subgraph and random walk sampling), $C$ will
have nonzero off-diagonal elements. Recall, however, that these
off-diagonal elements involved higher-order properties of the graph, in
the sense of summarizing even more structure than the degree
distribution we seek to estimate. Accordingly, it is unrealistic to
think to incorporate this information into our estimation strategy. We
instead focus on the diagonal elements of $C$.

We approximate the covariance matrix $C$ with a diagonal matrix of the form
%
\begin{equation}\label{eq:C.hat}
\hat{C}=\operatorname{diag}\bigl(\mathbf{N}_{\mathrm{smooth}}^*\bigr)+\delta I.
\end{equation}
The first term is a diagonal matrix with the diagonal entries equal to
a smoothed version of the observed degree vector. The arguments in
Section~\ref{sec:noise} suggest the merit of an approximate Poisson
variance for the diagonal elements of $C$, which in principle means
using $E[\mathbf{N}^*]=P\mathbf{N}$. Necessarily lacking this, it is
tempting to plug in the observed degree counts $\mathbf{N}^*$, but we
have found smoothing to offer noticeable improvement, as the noise in
the observations can be substantial. The discrete nature of $\mathbf{N}^*$ requires our using a smoothing method different from the
nonparametric methods used with continuous data. Here we employ the
kernel-smoothing method of \citet{dong1994construction}, which extends
the ideas in \citet{hall1987smoothing}, using an Epanechnikov kernel
with boundary correction, and least square cross-validation for
choosing an effective integer bandwidth.

To perform the weighted optimization in (\ref{eq:opt}), our proxy for
the covariance matrix $C$ must be positive definite. However, some of
the diagonal entries in the matrix $\operatorname{diag}(\mathbf{N}_{\mathrm{smooth}})$
typically are zero or close to zero. We adopt a standard strategy to
remedy this, by adding a small value $\delta$ to the diagonal elements.
We offer guidance on the choice of $\delta$ in the context simulation
and application in Sections~\ref{sec:simul} and~\ref{sec:appl}.

\section{Simulation study}
\label{sec:simul}

In this section we present a simulation study conducted to assess the
performance of the method we proposed in Section~\ref{sec:solve}, on
networks simulated from various random graph models. We also will look
at the effect of several factors (i.e., total number of vertices,
density and sampling rate) on the accuracy of the estimators. 

\subsection{Design}
There are several parameters that need to be chosen with some care.
Here we list them and discuss the conventions we applied:
\begin{itemize}
\item $\mathbf{b}$:   The random vector $\mathbf{b}$ must have
zero mean, covariance matrix $I$ and bounded higher order moments; here
we use a multivariate normal, that is, $\mathbf{b} \sim N (0,I)$.

\item $\varepsilon$:   In principle, the value $\varepsilon$ should be
small enough to approximate the notion of tending to zero, but not so
small as to induce floating point errors of an undesirable magnitude in
computing $f_{\lambda}(\mathbf{y}+\varepsilon\mathbf{b})$. In practice,
similar to the experience of~\citet{ramani2008monte}, we have witnessed
the method to be robust to choice of this parameter, even over several
orders of magnitude. In the following simulations, we use $\varepsilon= 0.1$.

\item $K$:   Small $K$ gives a noisy $\mathrm{WMSE}$ curve. As $K$
increases, we get a clearer shape for the $\mathrm{WMSE}$ curve and the
resulting estimate is more accurate. However, a~larger $K$ has bigger
computation cost. We have had good results using $K=100$.

\item $M$:   The maximum degree $M$ is set to be 1.1 times the true
maximum degree of the true graph in our simulations, to relax the
restriction of a known maximum degree.

\item $\delta$:   The parameter $\delta$ must be big enough to
make the optimization stable, but not so big as to swamp the
contribution of $\operatorname{diag}(\mathbf{N}_{\mathrm{smooth}})$ in (\ref{eq:C.hat}).
In these simulations, in order to make the results comparable across
different settings, we choose $\delta$ to make the condition number of
the approximate covariance matrix $\hat{C}$ the same, equal to $20$.


\item $\lambda$:  The range of $\lambda$ being considered in
finding the optimal $\lambda$ includes the true optimal $\lambda$ and
values of three magnitudes above and below the true $\lambda$.
\end{itemize}

To compare the estimated with the true degree distribution, we use the
Kolmogorov--Smirnov D-statistic, which has been used widely in the
literature on sampling of social media networks to illustrate the
accuracy of various sampling methods [e.g., \citet{leskovec2006sampling}, \citet{hubler2008metropolis},
\citet{ahmed2011network}]. The
statistic corresponds to the maximum difference between the two
cumulative distribution functions $F_1$ and $F_2$, that is, $D=\max_x\{
|F_1(x)-F_2(x)|\}$, and ranges from zero to one.

\subsection{Results}
\begin{figure}

\includegraphics{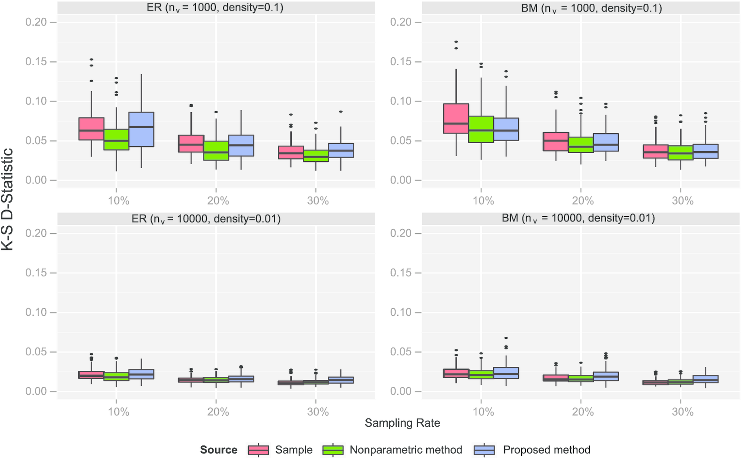}

\caption{Simulation results for ego-centric sampling. Error measured
by K--S D-statistic. For each sampling rate, the three boxes from left
to right represent K--S D-statistic comparing the true degree
distribution with (left) sample degree distribution, (middle) estimated
degree distribution using the nonparametric method and (right)
estimated degree distribution using the proposed method. (Online
versions of the figure are in color.)}
\label{fig:sim_ego}
\end{figure}

\begin{figure}

\includegraphics{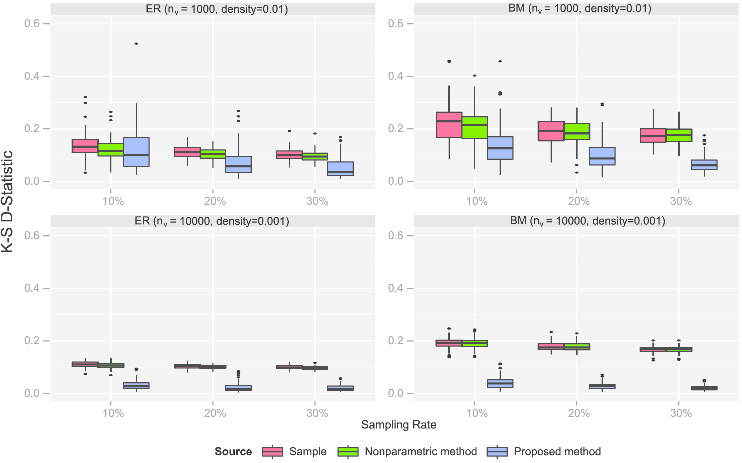}

\caption{Simulation results for one-wave snowball sampling. Error
measured by K--S D-statistic. For each sampling rate, the three boxes
from left to right represent K--S D-statistic comparing the true degree
distribution with (left) sample degree distribution, (middle) estimated
degree distribution using the nonparametric method and (right)
estimated degree distribution using the proposed method. (Online
versions of the figure are in color.)}
\label{fig:sim_snow}
\end{figure}

\begin{figure}

\includegraphics{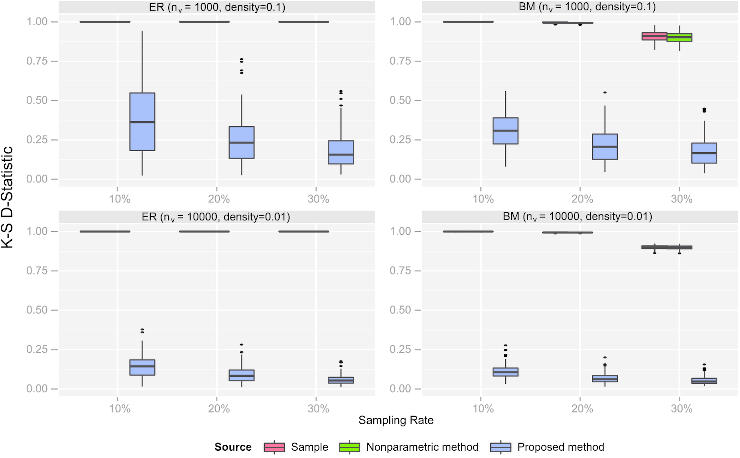}

\caption{Simulation results for induced subgraph sampling. Error
measured by K--S D-statistic. For each sampling rate, the three boxes
from left to right represent K--S D-statistic comparing the true degree
distribution with (left) sample degree distribution, (middle) estimated
degree distribution using the nonparametric method and (right)
estimated degree distribution using the proposed method. (Online
versions of the figure are in color.) (Note: Only the performance of
the proposed estimator $\hat{N}$ avoids the extremes of 1.0 in most cases.)}
\label{fig:sim_ind}
\end{figure}

Results of our simulation study are shown in Figures~\ref{fig:sim_ego}--\ref{fig:sim_ind}, for ego-centric, induced subgraph and one-wave
snowball sampling, respectively. Each box plot represents the
D-statistics computed from $100$ trials, that is, based on $100$
samples drawn from the underlying networks. Two types of networks are
studied: those from the Erd\"{o}s--R\'{e}nyi model and those from a
block model with two blocks. These are two basic models commonly used
in network studies [e.g., \citet{kolaczyk2009statistical}, Chapter~6].
In the Erd\"{o}s--R\'{e}nyi model, edges are randomly assigned to each
pair of vertices with a given probability, that is, the expected
density of the network. For the block model, each of the two blocks
itself is an Erd\"{o}s--R\'{e}nyi model. In addition, vertices from
different blocks are connected with some probability too. In the
simulation, edge probabilities for within the two blocks and between
blocks satisfy a ratio of $6\dvtx 2\dvtx 1$. For each of the two models, we let the
density and $n_v$ change but fix the average degree to be approximately
equal. In ego-centric and induced subgraph sampling, $n_v \times
\mathit{density} = 100$. In one-wave snowball sampling, we make $n_v \times
\mathit{density} = 10$. We have to use a lower average degree in one-wave
snowball sampling to avoid including all vertices of the true network
into the sample. In addition, the sampling rates of $10\%$, $20\%$ and
$30\%$ for one-wave snowball sampling indicate the percentage of the
total vertices of the two sequential selections.

Notice that the scale of Figure~\ref{fig:sim_ego} is from $0$ to $0.2$,
much smaller than that of Figure~\ref{fig:sim_snow} which is from $0$
to $0.6$, and Figure~\ref{fig:sim_ind} which is from $0$ to $1$. The
scales of the K--S D-statistics match the difficulty of the inverse
problems they come from, with ego-centric sampling yielding an easier
problem than one-wave snowball and induced subgraph sampling, as was
discussed in Section~\ref{sec:char}. We compare the estimated degree
distributions from our method with the sample degree distributions and
the estimates from a standard kernel-smoothing method [\citet{dong1994construction}] described in Section~\ref{sec33}. Only in the case of
ego-centric sampling, the sample degree distribution and the
kernel-smoothing method are competitive with our method. For one-wave
snowball and induced subgraph sampling, our method yields much better
results than the sample and kernel-smoothing method. This is to be
expected, of course, since the kernel-smoothing method does not account
for the underlying inverse problem.

In Figures~\ref{fig:sim_ego}--\ref{fig:sim_ind}, the performance in the
second row is better than the performance in the first row in general.
That is, performance improves with larger networks of lower density,
given fixed average degree. There are three reasons for this
phenomenon. First, in the standard Erd\"{o}s--R\'{e}nyi model, as $n_v$
grows to infinity and the density shrinks to zero, while the average
degree is fixed, the degree distribution becomes smoother and reaches a
Poisson distribution in the limit. Second, as density shrinks and $n_v$
grows, the normal/Poisson approximation of $N_k^*$, for $k=0,1,\ldots,
M$, is better. And, in turn, the approximation of covariance matrix $C$
is more accurate.

Comparing Erd\"{o}s--R\'{e}nyi and the block model under the induced
subgraph sampling (Figure~\ref{fig:sim_ind}), the block model has a
broader range of degrees than the Erd\"{o}s--R\'{e}nyi model at any
given choice of our other simulation parameters. In (\ref{eq:pois_appro.ind}), for each $k$, the indicator function involving $u
\in V$ with higher $d_u$ has lower probability of being equal to $1$.
Thus, a better Poisson approximation of $N_k^*$ and a more accurate
approximation of $C$ occur under the block model. A power-law network
has an even broader degree distribution. For the same reasons,
therefore, we expect the estimators for the power-law like networks in
the applications of Section~\ref{sec:appl} to perform similarly well. However, the
results for Erd\"{o}s--R\'{e}nyi and the block model are quite close in
Figures~\ref{fig:sim_ego} and~\ref{fig:sim_snow}. This is because
only the vertex with degree $k$ in the true network can possibly
contribute to degree $k$ under ego-centric and one-wave snowball sampling.

Three sampling rates are studied: 10\%, 20\%, and 30\%. Our results
show that there is less accuracy for smaller sampling rate, as is to be
expected. In the literature on Internet community monitoring, 30\%
sampling rates have been suggested as reasonable for preserving network
properties to a reasonable accuracy [\citet{leskovec2006sampling}]. In
our results, we see that our estimators of degree distribution perform
fairly well based on as low as a 10\% sampling rate.
%

%

\section{Applications}
\label{sec:appl}

The cost of any sampling strategy varies with the structure of the
network and the protocol. As we have remarked, sampling is of
particular interest in the context of online social networks. In online
social networks where each user is assigned a unique user id, it is a
common practice to select a set of users by querying a set of randomly
generated user id's [\citet{ribeiro2010estimating}]. Thus, our induced
subgraph sampling can be applied there. In this section, we use our
degree distribution estimation method on data from three online social
networks: Friendster, Orkut and LiveJournal. These data are available
on the SNAP (Stanford Network Analysis Project) website. In the
following we present our estimates of various degree distributions from
these online social networks. In addition, we show how these degree
distributions help us to gain insight about the epidemic thresholds of
these networks, which is relevant to the concept of social influence,
spread of rumors and viral marketing.

\subsection{Estimating degree distributions from online social networks}
\label{sec:degree}
It is now well understood that large-scale, real-world networks
frequently have heavy-tailed degree distributions.
\citet{stumpf2005sampling} proved analytically that for a network with
an exact power-law degree distribution, although its sampled network
under our sampling method [induced Subgraph sampling with
$\operatorname{Bernoulli}(p)$ for selecting vertices] is not an exact power-law network, the
degree distribution for large enough degrees is power law and has the
same exponent with the true network. In reality, however, most networks
with heavy-tailed degree distribution will \textit{not} have an exact
power law. Many, for example, exhibit exponential-like deviation from a
power law after some cutoff. As a result, the result of \citet{stumpf2005sampling} does not hold in such situations and estimation is
therefore still of fundamental interest.

In addition, the full Friendster, Orkut and LiveJournal networks
arguably are of less interest here, being a rather coarse-grained
aggregation of much finer-scale social interactions. Accordingly, we
focus instead on the estimation of degree distributions for subnetworks
corresponding to certain communities within these networks. In these
online social networks, users create functional groups that others can
join, based on, for example, topics, shared interests and hobbies, or
geographical regions. In our application, we use ground-truth
communities established by \citet{yang2012defining}. For example, these
authors found that LiveJournal categorizes social groups into the
categories of ``culture, entertainment, expression, fandom, gaming,
life/style, life/support, sports, student life and technology'' [\citet{yang2012defining}]. It is the degree distributions for subnetworks
corresponding to collections of ground-truth communities such as these
that we estimate here.
%
\begin{figure}

\includegraphics{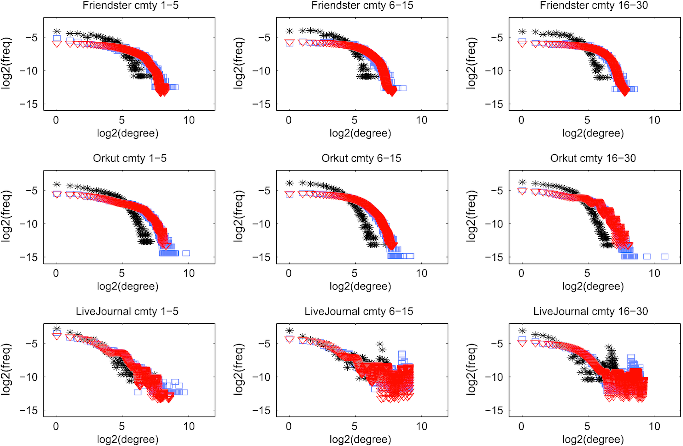}

\caption{Estimating degree distributions of communities from
Friendster, Orkut and LiveJournal.
Squares represent the true degree distributions, stars represent
the sample degree distributions,
and triangles represent the estimated degree distributions.
Sampling
rate $=$ 30\%. Points which correspond to a density${}<10^{-4}$ are
eliminated from the plot. (Online versions of the figure are in color.)}
\label{fig: appl_fig}1
\end{figure}


Figure~\ref{fig: appl_fig} gives an example of the estimators. The
first row is for three subnetworks from Friendster. Communities are
ordered according to the number of users in them. In the top left
subplot, vertices from the top 5 communities form an induced subnetwork
for which the degree distribution is to be estimated. Then Bernoulli
sampling of vertices with $30\%$ sampling rate is performed on this
subnetwork, and our estimation method is applied. Similarly, the true
network in the top middle plot is induced by the top 6--15 communities,
and in the top right plot the true network is induced by the top 16--30
communities. The second row and the third row show estimates of Orkut
and LiveJournal, respectively. Examination of these plots shows that,
while the sampled degree distribution can be quite off from the truth,
particularly in the case of the Friendster and Orkut networks,
correction for sampling using our proposed methodology results in
estimates that are nearly indistinguishable by eye from the true degree
distributions.

In Table~\ref{table:cmty} the median and inter-quartile range are
computed based on the application of our estimator to 20 samples. The
estimated degree distribution greatly improves over the degree
distribution of the sample, as measured by the K--S D-statistic. In
fact, the improvement in accuracy is by an order of magnitude, with the
values of the D-statistic produced by our estimator being on the same
order of magnitude as the best results in our simulation study.
\begin{table}
\tabcolsep=0pt
\tablewidth=\textwidth
\caption{Network communities summary. Each median and inter-quartile
range is computed based on the application of our estimator to 20 samples}
\label{table:cmty}
\begin{tabular*}{\tablewidth}{@{\extracolsep{4in minus 4in}}lcccccccc@{}}
\hline
&&&&& \multicolumn{2}{c}{\textbf{Sample}} & \multicolumn{2}{c@{}}{\textbf{Estimator}}\\
&&\multicolumn{1}{c}{\multirow{2}{40pt}[-8pt]{\centering{\textbf{Numbers of vertices}}}}&
\multicolumn{1}{c}{\multirow{2}{35pt}[-8pt]{\centering{\textbf{Numbers of edges}}}}& &
\multicolumn{2}{c}{\textbf{D-statistic}}&
\multicolumn{2}{c@{}}{\textbf{D-statistic}}\\[-4pt]
&&&&&\multicolumn{2}{l}{\hrulefill} & \multicolumn{2}{l@{}}{\hrulefill}\\
\textbf{Net}&\textbf{cmty}&&&\textbf{dmax}&\textbf{Median}&\textbf{IQR}&\textbf{Median}&\textbf{IQR}\\
\hline
&\phantom{0}1--5\phantom{0}&\phantom{0,}5748&163{,}888& \phantom{0}494&0.4242&0.0196&0.0221&0.0080\\
Friendster&\phantom{0}6--15&\phantom{0,}6385&131{,}875&\phantom{0}383& 0.4521&0.0164&0.0187&0.0107\\
&16--30&\phantom{0,}7097&162{,}616&\phantom{0}357&0.4813&0.0211& 0.0143&0.0161\\
&\phantom{0}1--5\phantom{0}&22{,}059&689{,}659&\phantom{0}895& 0.4092&0.0145&0.0134&0.0073\\[3pt]
Orkut&\phantom{0}6--15&29{,}681&591{,}448&\phantom{0}578& 0.4322&0.0129& 0.0099&0.0059\\
&16--30&31{,}018&619{,}909&1779& 0.4324&0.0068&0.0175&0.0076\\
&\phantom{0}1--5\phantom{0}&\phantom{0,}5131& \phantom{0}85{,}419&\phantom{0}801&0.3018&0.0285&0.0430&0.0258\\[3pt]
LiveJournal&\phantom{0}6--15&\phantom{0,}3757&219{,}193&\phantom{0}547&0.2678&0.0153&0.0558& 0.0105\\
&16--30&\phantom{0,}4591&228{,}633&\phantom{0}512&0.2941&0.0137&0.0643& 0.0404\\
\hline
\end{tabular*}
\end{table}

In summary, our method of estimating the degree distribution from
sampled networks clearly can offer substantial advantages over raw
measured networks in monitoring the degree distribution of the
communities in online social networks. This provides a powerful
additional motivation for using sampling in these contexts.

\subsection{Characterizing epidemic spread}
\label{sec:epidemic}
In this subsection we are going to show how recovery of the degree
distribution---as a fundamental object---helps for monitoring other
socially pertinent questions, for example, characterizing epidemic
spread on networks.

As has been shown by various authors [e.g.,
\citet{bailey1975mathematical,daley1999epidemic,kephart1991directed,pastor2001epidemic}], an epidemic threshold $\tau_c$ exists in a virus
spread in networks. Under a standard Susceptible--Infected--Susceptible
(SIS) model, let the infection rate be $\beta$ and the curing rate be
$\delta$. If the effective spreading rate $\tau=(\beta/\delta)>\tau_c$,
the virus persists and a nonzero fraction of the nodes are infected,
whereas for $\tau\leq\tau_c$ the epidemic dies out. This threshold is
shown to equal the inverse of the largest eigenvalue $\lambda_1$ of the
network's adjacency matrix in \citet{van2009virus}.

The degree distribution of a network can be used to get bounds for the
largest eigenvalue $\lambda_1$ of the adjacency matrix, and thus bounds
for $1/\lambda_1$. Let $M_1$ be the first raw moment of the degree
distribution, that is, the average degree, $M_2$ be the second raw
moment of the degree distribution, $n_e=|E|$ be the number of total
edges, and $U=(2*n_e(n_v-1)/n_v)^{1/2}$. Then we have the following
relationship:
%
\begin{equation}
M_1\leq\sqrt{M_2} \leq\lambda_1 \leq U.
\end{equation}
The proof of the first two inequalities can be found in \citet
{van2011graph}, and the third (upper bound) can be found in \citet
{lovasz1993combinatorial}. Thus, we have the bounds for the epidemic
threshold $\tau_c$,
%
\begin{equation}
1/U \leq\tau_c \leq \frac{1}{\sqrt M_2} \leq\frac{1}{M_1}.
\end{equation}

\begin{figure}

\includegraphics{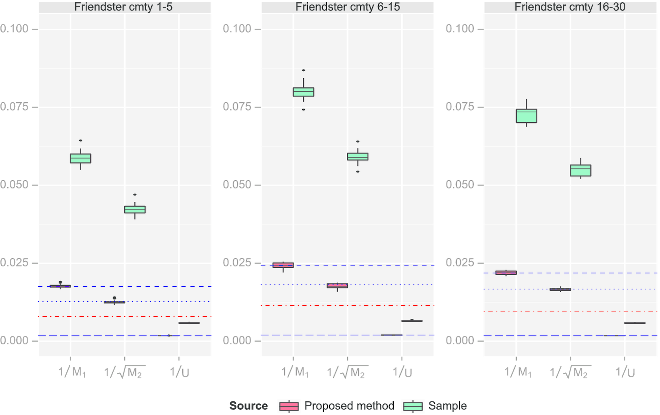}

\caption{Bounds for the epidemic spreads of Friendster networks, each
box is estimated based on~20 samples, four horizontal lines are the
true values for $\frac{1}{M_1}$, $\frac{1}{\sqrt{M_2}}$, $\frac
{1}{\lambda_1}$ and $\frac{1}{U}$ from top to bottom. For each bound,
the two boxes from left to right correspond to the estimated value
using (left) the proposed method and (right) the sample degree
distribution. (Online versions of the figure are in color.)}
\label{fig:SpectralFriendster}
\end{figure}

\begin{figure}

\includegraphics{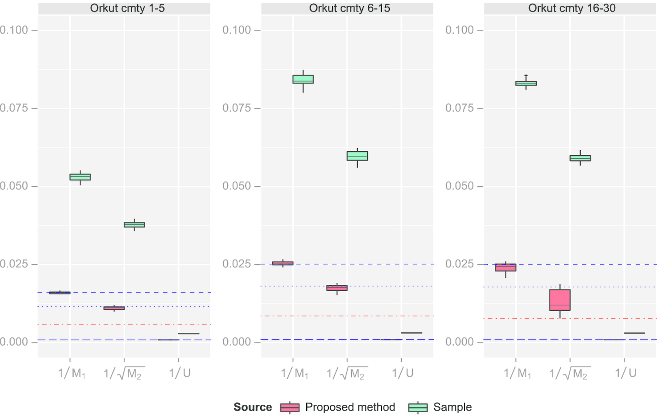}

\caption{Bounds for the epidemic spreads of Orkut networks, each box
is estimated based on 20 samples, four horizontal lines are the true
values for $\frac{1}{M_1}$, $\frac{1}{\sqrt{M_2}}$, $\frac{1}{\lambda
_1}$ and $\frac{1}{U}$ from\vspace*{1pt} top to bottom. For each bound, the two
boxes from left to right correspond to the estimated value using (left)
the proposed method and (right) the sample degree distribution. (Online
versions of the figure are in color.)}
\label{fig:SpectralOrkut}
\end{figure}

\begin{figure}

\includegraphics{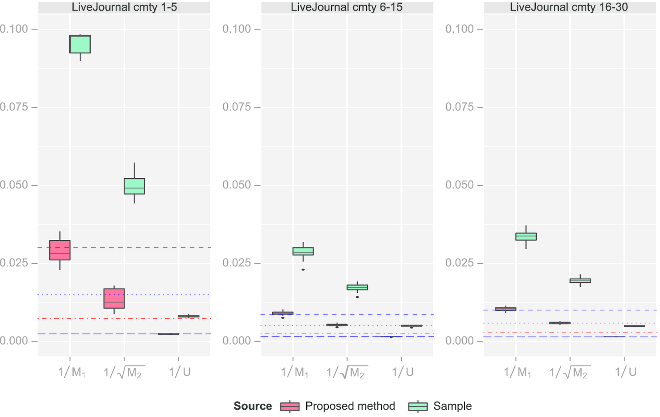}

\caption{Bounds for the epidemic spreads of LiveJournal networks,
each box is estimated based on~20 samples, four horizontal lines are
the true values for $\frac{1}{M_1}$, $\frac{1}{\sqrt{M_2}}$, $\frac
{1}{\lambda_1}$ and $\frac{1}{U}$ from top to bottom. For each bound,
the two boxes from left to right correspond to the estimated value
using (left) the proposed method and (right) the sample degree
distribution. (Online versions of the figure are in color.)}
\label{fig:SpectralLiveJournal}
\end{figure}

Figures~\ref{fig:SpectralFriendster}--\ref{fig:SpectralLiveJournal}
show the bounds obtained from the estimated degree distribution and
those obtained from the original sample degree distribution. The
networks used are the online social networks described in Section~\ref{sec:degree}. It can be seen from Figures~\ref{fig:SpectralFriendster}--\ref{fig:SpectralLiveJournal} that our method estimates the bounds
with high accuracy, whereas the bounds using the sampled data are way off.

Since our estimator successfully recovers the degree distribution of
the online social networks, the epidemic threshold (the inverse of the
spectral radius) of the network can be successfully bounded by
functions of our estimates. This has important implications in
practical applications. For example, in viral marketing, the epidemic
threshold relates to how hard a company's marketing force needs to
work, that is, it is necessary for them to make the effective spreading
rate $\tau$ as large as $1/U$, and sufficient to make $\tau$ as large
as $\frac{1}{\sqrt M_2}$, in order to make a product's advertisement
remembered by people in the network.

\section{Discussion}
\label{sec:disc}

The problem of estimating the degree distribution of a network from a
sampled subnetwork was first posed by Ove Frank in his 1971 Ph.D.
dissertation [\citet{frank1971statistical}]. In the ensuing years, the
problem appears to have received very little attention, likely in no
small part to its apparent difficulty. Here we recast the original
problem as a linear inverse problem. We have demonstrated that, in so
doing, it is possible to obtain substantial insight into the inherent
difficulty of the problem---in terms of the operator corresponding to
the sampling, the nature of the ``noise'' induced by the sampling and
the manner in which the two interact. Leveraging this insight, we have
proposed a penalized, generalized least squares estimator, with
positivity constraints, that solves our linear inverse problem. The
choice of smoothing parameter is nontrivial in this context and we
offer a Monte Carlo approach to optimizing a generalized SURE criterion
as an effective option. Finally, our simulations and application to
online social media networks show that the methodology can perform
quite well under a variety of choices of network topology---even under
sampling rates as low as 10\%.

There are a number of directions upon which to build from the work we
present here. The assumptions discussed in Section~\ref{sec24} could be
relaxed, for example, to include observation errors, to incorporate
estimates of possible unknown parameters in the matrix $P$, or to focus
on matrices $P$ which depend on the network $G$ itself. In this case, a
model-based framework is likely necessary, and for that it would be
natural to try to integrate our framework with the work of~\citet
{handcock2010modeling}. Finally, another interesting direction would be
developing methods for correcting the sampling bias of the degree
distribution under more complex adaptive designs.


\begin{appendix}
\section{Eigenvalue decomposition}\label{appen:eigen}
%
\begin{thm}
\label{thm:eigen}
Let $P=P_{\mathrm{ind}}=\tilde{U}\Lambda\tilde{U}^{-1}$, where $\Lambda
=\operatorname{diag}(\lambda_1,\ldots,\lambda_{M+1})$ is a diagonal matrix and
$\tilde{U}=(\tilde{\mathbf{u}}_1,\tilde{\mathbf{u}}_2,\ldots,
\tilde{\mathbf{u}}_{M+1})$ is a nonsingular matrix. Then the $k$th eigenvalue
$\lambda_k$ and eigenvector $\tilde{u}_k$ of $P$ are
%
\begin{eqnarray}
\lambda_k &=& p^k,
\\
\tilde{\mathbf{u}}_k(j) &= &
\cases{
\ds (-1)^{k-j} \pmatrix{k-1 \cr j-1}, &$\quad \mbox{for } 1\leq j \leq k$,
\vspace*{3pt}\cr
0,  &\quad $\mbox{for } k<j \leq M+1$.}
\end{eqnarray}
\end{thm}
\begin{pf}
 We will prove this theorem by
induction. In the case that $P$ is a $2$ by~$2$ matrix,
%
\begin{equation}
P = \left[
\matrix{ p & pq
\vspace*{2pt}\cr
0 & p^2}\right].
\end{equation}

It's easy to show that
%
\begin{equation}
\tilde{U}=\left[
\matrix{ 1 & 0\vspace*{2pt}\cr
-1 & 0}\right].
\end{equation}

The theorem is true if $P$ is a $2$ by $2$ matrix. Suppose it is true
when $P$ is a $k-1$ by $k-1$ matrix, then in the case that $P$ is $k$
by $k$,
%
\begin{equation}
\qquad {\fontsize{10.3}{11.3}\selectfont{P= \left[\matrix{ \ddots& & & &
\vspace*{3pt}\cr
&p^{k-3}& \pmatrix{k-3\cr 1}p^{k-3}q &\pmatrix{k-2\cr 2}p^{k-3}q^2
&\pmatrix{k-1\cr 3}p^{k-3}q^3
\vspace*{4pt}\cr
&0&p^{k-2}&\pmatrix{k-2 \cr 1}p^{k-2}q &\pmatrix{k-1\cr 2}p^{k-2}q^2
\vspace*{4pt}\cr
&0&0&p^{k-1}&\pmatrix{k-1\cr 1}p^{k-1}q
\vspace*{4pt}\cr
&0& 0&0&p^k}\right].}}\hspace*{-10pt}
\end{equation}

Because of the upper-triangular nature of the matrix, the first $k-1$
entries in each of the first $k-1$ eigenvectors are the same as in the
case that $P$ is $k-1$ by $k-1$, and the $k$th entry is filled with zero.

For eigenvalue $\lambda_k=p_k$, let $\mathbf{x}=(x_1,x_2,\ldots,
x_k)^T$ and $x_k=1$ be the solution of the eigenvalue equation
%
%
\begin{eqnarray}
&& (P-\lambda_k I)\mathbf{x}\nonumber\\
&& \qquad = {\fontsize{9.5}{10.5}\selectfont{\left[
\matrix{ \ddots& & & &
\vspace*{4pt}\cr
&p^{k-3}\bigl(1-p^3\bigr)&\pmatrix{k-3\cr 1}p^{k-3}q
&\pmatrix{k-2\cr 2}p^{k-3}q^2 &\pmatrix{k-1\cr 3}p^{k-3}q^3
\vspace*{4pt}\cr
&0&p^{k-2}\bigl(1-p^2\bigr)&\pmatrix{k-2 \cr 1}p^{k-2}q
&\pmatrix{k-1\cr 2}p^{k-2}q^2
\vspace*{4pt}\cr
&0&0&p^{k-1}(1-p)&\pmatrix{k-1\cr 1}p^{k-1}q
\vspace*{4pt}\cr
&0& 0&0&0}\right]}}
\mathbf{x}\hspace*{-6pt}\\
&&\qquad =0.\nonumber
\end{eqnarray}

The equation at the $(k-1)$th row is
%
\begin{equation}
p^{k-1}(1-p)x_k+\pmatrix{k-1\cr 1}p^{k-1}q
x_k=0.
\end{equation}

We solve for $x_{k-1}$,
%
\begin{equation}
x_{k-1}=\frac{{k-1\choose 1}p^{k-1}q
}{p^{k-1}(1-p)}=-\pmatrix{k-1\cr 1}.
\end{equation}

Assuming $x_{k-i}=(-1)^i{k-1 \choose i}$, for $i=0,1,\ldots,n-1$, we solve
for $x_{k-n}$ from the equation at the ($k-n$)th row:
%
\begin{eqnarray}
-p^{k-n}\bigl(1-p^n\bigr)x_{k-n}&=&\pmatrix{k-n
\cr 1}p^{k-n}q x_{k-(n-1)}
\nonumber
\\
&&{}+\pmatrix{k-(n-1) \cr 2}p^{k-n}q^2x_{k-(n-1)}+ \cdots
\\
&&{}+\pmatrix{k-2\cr n-1}p^{k-n}q^{n-1}x_{k-1}+\pmatrix{k-1\cr
n}p^{k-n}q^n x_k.\nonumber
\end{eqnarray}

Simplifying the above equation, we have
%
\begin{eqnarray}
&& -\bigl(1-p^n\bigr)x_{k-n}\nonumber\\
&&\qquad= \pmatrix{k-n \cr 1} \pmatrix{k-1 \cr
n-1}(-1)^{n-1}q
\nonumber
\\
&&\qquad\quad{}+\pmatrix{k-(n-1) \cr 2} \pmatrix{k-1 \cr n-2}(-1)^{n-2}q^2+ \cdots
\nonumber
\\
&&\qquad\quad{}+\pmatrix{k-2\cr n-1} \pmatrix{k-1 \cr 1}(-1)^1q^{n-1}+\pmatrix{k-1\cr
n}(-1)^0q^n
\nonumber
\\[-8pt]
\\[-8pt]
\nonumber
&&\qquad=(-1)^n\pmatrix{k-1 \cr n} \\
\nonumber
&& \qquad\quad{}\times\biggl[ \pmatrix{n \cr 1}(-q)+\pmatrix{n \cr
2}(-q)^2+\cdots+\pmatrix{n \cr 1} (-q)^{n-1}+\pmatrix{n \cr 0}(-q)^n
\biggr]
\\
&&\qquad=(-1)^n \pmatrix{k-1 \cr  n}\bigl[(1-q)^n-1\bigr]
\nonumber
\\
&&\qquad=(-1)^n \pmatrix{k-1 \cr n} \bigl(p^n-1\bigr).\nonumber
\end{eqnarray}

Finally,
%
\begin{equation}
x_{k-n}=(-1)^{n}\pmatrix{k-1 \cr n}.
\end{equation}

Therefore, the entries in the $k$th eigenvector are
%
\begin{equation}
\tilde{\mathbf{u}}_k(j) = \cases{
\ds (-1)^{k-j} \pmatrix{k-1 \cr j-1},  &\quad$\mbox{for } 1\leq j \leq k$,
\vspace*{3pt}\cr
0,  & \quad$\mbox{for } k<j \leq M+1$.}
\end{equation}

The theorem is true for $k$ by $k$ matrix $P$.
\end{pf}

\section{Poisson approximation}\label{appen:pois}
Here we give a proof of the Poisson approximation of the cumulative
degree vectors, under one-wave snowball sampling and induced subgraph
sampling with $\operatorname{Bernoulli}(p)$ for selecting edges. The arguments for
both designs are nearly identical, and so we present them together.

\begin{thm}
\label{thm:Poisson}
Assume $G^*$ is produced by induced subgraph sampling with Bernoulli
sampling to select $S$. Let
%
\begin{equation}\label{eq:cum.cnt}
\tilde{N}_k^* = \sum_{r=k}^{M}
N_r^* = \sum_{v} I\bigl\{v\in S,
d_v^*\ge k\bigr\}
\end{equation}
be the number of vertices of degree $k$ or larger in $G^*$. Let
%
\begin{equation}\label{eq:tau}
\lambda_k=E\bigl(\tilde N_k^*\bigr) = \sum
_{v:d_v\ge k} \pi_{k,v},
\end{equation}
where
%
\begin{equation}\label{eq:pi}
\pi_{k,v} = P \bigl( v\in S, d_v^*\ge k \bigr).
\end{equation}
Then
%
\begin{equation}\label{eq:CSbnd}
\hspace*{6pt}\operatorname{dist}_{\mathrm{TV}} \bigl(\mathcal{L}\bigl(\tilde N_k^*\bigr),
\operatorname{Po}(\lambda_k) \bigr) \le \frac{1-e^{-\lambda_k}}{\lambda_k} \biggl[ \operatorname{Var}
\bigl(\tilde{N}_k^*\bigr) - \lambda_k + 2 \sum
_{v\dvtx d_v \ge k} \pi_{k,v}^2 \biggr],
\end{equation}
where $\operatorname{dist}_{\mathrm{TV}}$ indicates the total-variation distance between
its arguments, $\mathcal{L}$ means ``law of,'' and
$\operatorname{Po}(\lambda_k)$ is a Poisson random variable with intensity $\lambda_k$.
\end{thm}

\begin{pf}
We sketch the proof briefly
here. Without loss of generality, (partially) order the vertices $\{
v_1,\ldots,v_{n_v}\}$ by (non)decreasing degree. Associate a binary
random vector $(X_1,\ldots,X_{n_v})$ with the vertices, where the
elements are independent Bernoulli random variables with parameter $p$.
So $\mathbf{X}$ represents the selection of vertices for inclusion in
$S$ in the case of induced subgraph sampling and the initial selection
of vertices in the case of snowball sampling. Now let $I_{v,k}$ be an
indicator random variable, which is one if $v\in S$ and $d_v^*\ge k$.
Then the variables $I_{v,k}$ are so-called ``increasing functions'' of
realizations of $X$. So Corollary~2.E.1, page 28, of \textit{Poisson
Approximation}, by Barbour and colleagues, yields our result.

In more detail, there are two key observations to be made. First, we
need the $I_{v,k}$ to be increasing functions.
This induces positive correlation among these indicator variables and
it makes a general Chen--Stein bound become much cleaner, as in our
theorem, in that it can be expressed explicitly in terms of means and
variances. Partial ordering means that if we let $\mathbf{x}$ and
$\mathbf{y}$ be two possible realizations of $\mathbf{X}$, then $\mathbf
{x}\le\mathbf{y}$ if and only if $x_i\le y_i$ for all $i$. And a
function $f$ is increasing if $f(\mathbf{x})\le f(\mathbf{y})$ whenever
$\mathbf{x}\le\mathbf{y}$. For $\mathbf{x}$ to be less than or equal
to $\mathbf{y}$, it suffices to think of what happens simply when a new
vertex enters the sample $S$. One element of $\mathbf{x}$ will change
from a zero to a one, so $\mathbf{x}\le\mathbf{y}$. What happens to
$I_{v,k}$? If $v$ is a vertex that was already in $S$, under $x$, then
adding a vertex to the sample under $\mathbf{y}$ can either not change
or increase its degree. So $I_{v,k}(\mathbf{x})\le I_{v,k}(\mathbf
{y})$. On the other hand, if $v$ itself was the new vertex to enter $S$
under $y$, the same statement can be made.\vadjust{\goodbreak}

Second is the observation that elements of $\mathbf{X}$ are independent
in our setting, which is guaranteed by our assumption of Bernoulli
sampling. Taken together, these two things mean that Theorem~2.E holds
in Barbour  et al., that is, positive dependence. And so Corollary~2.E.1 holds and we have our result.
\end{pf}

\section{Conditions to use generalized SURE}\label{appen:sure}
\subsection{Weak differentiability of
\texorpdfstring{$f_{\lambda}(\mathbf{N}^*)$}
{flambda(N*)}}

Let's first ignore the nonnegativity constraints. Then \ref{eq:opt} becomes
%
\begin{eqnarray}
&& \mathop{\mbox{minimize}}_{N}  \qquad  \bigl(P
\mathbf{N}-\mathbf{N}^* \bigr)^TC^{-1} \bigl(P\mathbf {N}-
\mathbf{N}^* \bigr)+\lambda\cdot\operatorname{penalty}\bigl(\mathbf{N}^*\bigr)
\nonumber
\\[-8pt]
\label{eq:optwn}
\\[-8pt]
\nonumber
&& \mbox{subject to} \qquad \sum_{i=0}^M{N_i}=n_v.
\end{eqnarray}

The Lagrange function is
%
\begin{equation}
L=\bigl(\mathbf{N}^*-P\mathbf{N}\bigr)^T\bigl(\mathbf{N}^*-P
\mathbf{N}\bigr)+\lambda\mathbf {N}^T\Omega\mathbf{N}+\alpha\bigl(
\mathbf{1}^T\mathbf{N}-n_v\bigr).
\end{equation}

KKT conditions:
%
\begin{eqnarray}
\frac{dL}{d\mathbf{N}}&=&-2\mathbf{N}^{*T}C^{-1}P+2\mathbf
{N}^TP^TC^{-1}P+2\lambda\mathbf{N}^T
\Omega+\alpha\mathbf{1}^T=0,
\\
\mathbf{1}^T\mathbf{N} &=& n_v.
\end{eqnarray}

Then $\hat{N}$ is the solution of the following system:
%
\begin{equation}
\left[
\matrix{
P^TC^{-1}P+\lambda\Omega&
\tfrac{1}{2}\mathbf{1}
\vspace*{3pt}\cr
\mathbf{1}^T & 0
}\right]
\left[\matrix{ \mathbf{N}
\cr
\alpha}\right]
=
\left[
\matrix{2P^TC^{-1}
\mathbf{N}^*
\vspace*{2pt}\cr
n_v
}\right].
\end{equation}

Let $A=P^TC^{-1}P+\lambda\Omega$ and $B=\bigl[
{\fontsize{8.36}{9.36}\selectfont{\matrix{
A & \tfrac{1}{2}\mathbf{1}\vspace*{-0.5pt} \cr
\mathbf{1}^T & 0}
}}\bigr]$.
Since both $A$ and $\mathbf{1}^TA^{-1}\mathbf{1}$ are invertible for
sufficiently large $\lambda$, $B$ is invertible:
%
\begin{equation}\label{C6}
\hat{\mathbf{N}}=B ^{-1}P^TC^{-1}\mathbf{N}^*=
\sum_{i=0}^M d_i\bigl(\mathbf{u}_i^TC^{-1}\mathbf{N}^*\bigr)B^{-1}
\mathbf{v}_i.
\end{equation}

Thus, $\hat{\mathbf{N}}$ is a linear function of the observed $\mathbf
{N}^*$. In this case, $f_{\lambda}(\mathbf{N}^*)$ is differentiable
w.r.t. $\mathbf{N}^*$.

Adding nonnegativity constraints only gives nondifferentiable points
at the boundary, so the set of nondifferentiable points has measure zero.
$f_{\lambda}(\mathbf{N}^*)$ has a derivative almost everywhere.
$f_{\lambda}(\mathbf{N}^*)$ is weakly differentiable.

\subsection{\texorpdfstring{$E\{|f_{\lambda}(\mathbf{N}^*)|\}$}
{E\{|flambda(N*)|\}} is bounded}

Assuming $\mathbf{N}^*$ is Gaussian, since $f_{\lambda}(\mathbf{N}^*)$
is a linear function of $\mathbf{N}^*$ within the feasible set of
$\hat{\mathbf{N}}$, $f_{\lambda}(\mathbf{N}^*)$ is also Gaussian, thus
$E \{|f_{\lambda}(\mathbf{N}^*)|\}$ is bounded.
\end{appendix}

\section*{Acknowledgment}
This work was begun during the 2010--2011 Program on
Complex Networks at SAMSI.

%

%

%


\printaddresses
\end{document}